\documentclass[twoside]{ilcws10}
\usepackage[latin1]{inputenc}
\usepackage[dvips]{graphicx,epsfig,color}
\usepackage{wrapfig,rotating}
\usepackage{amssymb,amsmath,array}
\usepackage{hyperref}

\begin{document}
\title{
%%%%   Paper title goes here  %%%%%%%%%%%%%%
Semi-DHCAL software developments: Digitization and Display} %% 
%***********************************************************************
% AUTHORS INFORMATION AREA
%***********************************************************************
\author{Manqi Ruan
% Optional short acknowledgment: remove next line if non-needed
%\thanks{This is an optional funding source acknowledgment.}
\thanks{ On the behavior of SDHCAL collabration }
% DO NOT MODIFY THE FOLLOWING '\vspace' ARGUMENT
\vspace{.3cm}\\
% Addresses and institutions (remove "1- " in case of a single institution)
%1- First Author's Institution - Department \\
%Address of First Author's Institution - Country
Laboratoire Leprince-Ringuet (LLR), Ecole polytechnique \\
91128 Palaiseau, France
%% Remove the next three lines in case of a single institution
%\vspace{.1cm}\\
%2- Department of Physics, Indian Institute of Technology, \\
%208016 Kanpur, India
}
%%***********************************************************************
% END OF AUTHORS INFORMATION AREA
%***********************************************************************

\maketitle

\begin{abstract}

GRPC Semi-Digital HCAL is a solid option for the PFA oriented calorimetry of the International Linear Collider.
Together with the hardware, the software developments is progressing steadily. The stauts and plans for the
GRPC SDHCAL software development are presented, as well the first order digitization module for the GRPC and the 
display program DRUID (Display Root module Used for ILD) have been introduced.

\end{abstract}

\section{Introduction}

%The majority of interested physics channels in the ILC contain jets in their final states (i.e, the ZH events, $t\bar{t}$ Events), so the precise reconstruction of the jets four-momentum is of essential importance to the R \& D of the ILC detector. 
%The goal is to reach a jet energy resolution accuracy of $\delta E_{jet}=30\%\sqrt{E_{jet}}$, which is a factor of two better than what has been achieved in the previous experiment. %\cite{ALEPH}% 
%It's believed that the best jet energy resolution could be achieved by the Particle Flow Algorithm together with a calorimeter system with very high granularity.

The Semi-Digital Glass RPC technology is a solid option for the ILD detector, 
because the RPC is a well-understood technology with many advantages, such as high efficiency, homogeneous, robust and low cost;
as a gaseous detector, the RPC is almost free of neutron hits;
and most importantly, the semi-digital technology codes every channel with only two bits, allows a granularity as high as 1 channel per square cm with relatively low electronic cost.
%Good performance is expected with preliminary simulation study.
Good performance is expected in jet energy resolution. 

%\begin{figure}
%\begin{wrapfigure}{r}{1.0\columnwidth}
%\begin{center}
%\centerline{\includegraphics[width=1.0\columnwidth]{Prototypes.eps}}
%\caption{GRPC SDHCAL Prototypes: mini-DHCAL, square meter and ANR cubic meter}
%\label{Prototype}
%\end{center}
%\end{wrapfigure}
%\end{figure}

%\begin{wrapfigure}{r}{0.5\columnwidth}

%\begin{figure}
%\centerline{\includegraphics[width=0.8\columnwidth]{Druidshow.eps}}
%\caption{Druid: $e^{+}e^{-} \rightarrow HZ \rightarrow \tau\tau\mu\mu $ event at 230 GeV}
%\label{Druidshow}
%\end{wrapfigure}

The GRPC SDHCAL (GRPC Semi-Digital HCAL) group is an international collabration consists of more than ten institutes. 
%And we are in communication with the US DHCAL group.
The aim is to have a good understanding to the GRPC SDHCAL detector and to proof the GRPC SDHCAL is a feasible option to the ILC detector.

\begin{figure}[h]
\centerline{\includegraphics[width=1.0\columnwidth]{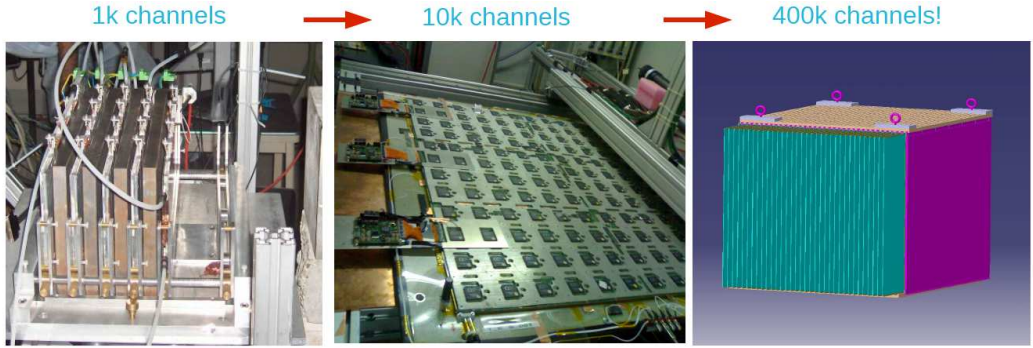}}
\caption{GRPC SDHCAL Prototypes: mini-DHCAL, square meter and ANR cubic meter}
\label{Prototype}
\end{figure}

In order to understand the detector performance and to optimize the experimental setting, prototypes (see Fig.~\ref{Prototype}) are constructed and tested with various options.
We have the so called mini-DHCAL with 1k channel, the square meter with 10k channel, and the ANR cubic meter with 400k channel is under constrtuction.
Lots of new technologies will be tested on the cubic meter, such as power pulsing, embedded electronics and self-supporting mechanics. 
Besides, the analysis of the test beam data will be used to validate the Geant 4 hadronic simulation.

% since we know there exist significant discrepancy between simulation \& data.

Now let's start from an introudction on the status and plan of GRPC Semi-DHCAL softwares. 

\subsection{Staus and plan for the GRPC SDHCAL softwares}

As well as for the hardware, the GRPC SDHCAL group has made great efforts on the software development. 

For the test beam experiment, a software chain (DAQ, reconstruction, analysis) based ROOT has been developed.
The data format will be updated to LCIO standard, and currently people are focusing on the R \& D of the new DAQ system.

A Geant 4 simulation package with the cubic meter prototype geometry has been developed,
and a valid concept for the ILD with a la Videau DHCAL Geometry and GRPC sensor layer is currently availalbe in Mokka, the standard ILC detector simulation software.
%and we plan to validate different options with ILD geometry in Mokka. 

The digitization is important for the GRPC SDHCAL, as we believed the semi-digital option could kept more useful information than the pure digital HCAL. 
A first order digitization module with the cosmic ray experimental input has been developed, which converts the shower particle energy deposition in each cell to the quantity of induced charge on the electrinic nodes. 
%The preliminary result shows that because of the large uncertainty in the RPC avanche development, we have large smearing on the induced charge spectrum for the low energy deposition,
%but statistically at high energy deposition, we have good linear correlation between this two quantity. 
%In other word, the induced charge could successfully represnent the energy deposition information at high energy deposition region, ensures better performance of Semi-DHCAL.
Preiliminary result is optimistic. 
The digitization module has been integrated into standard Marlin framework, and it will be upgraded with multiplicity and saturation correction. 

%Because the semi-DHCAL has much high granularity than the AHCAL, people expect better performance for the jet energy reconstruction. 
%And this is supported by some preliminary simulation study (ref..).

Because GRPC SDHCAL utilises totally different technology and detector geometry comparing to the AHCAL concept, 
its reconstruction algorithm need to be developed and optimized, which is regarded as one of the central task for SDHCAL software development. 
People are studying the hadronic shower development, aiming at develop an optimized reconstruction algorithm for the GRPC SDHCAL.
The idea is to integrate the SDHCAL reconstruction algorithm into the full Pandora reconstruction chain \cite{Pandora}, to achieve the best performance in jet energy resolution. 

A 3d display program DRUID based on root \cite{rootpage} TEve class ( developed for the LHC event display ) has been developed. 
It can visualize the information stored in standard ilc data file (slcio file) with various options and different detector geometries.
With DRUID, better understanding of the ILD/test beam event and reconstruction algorithm performance could be achieved.
%as DRUID can display Monte Carlo truth objects (MCParticle, simulated detector hits, etc) together with the reconstructed objects, 
%Druid could be used to analysis the performance of reconstruction algorithm. 
Some examples are given in the following section. 
 
Central MC generation with particle gun events and ILC benchmark physics processes is also proposed, 
which will server as a basis for the algorithm development, performance analysis and geometry optimization study. 

Now let's fouse on the software tools of digitization and display. 

\begin{figure}[h]
%\centerline{\includegraphics[width=0.95\columnwidth]{Cosmic.eps}}
\centering
\includegraphics[width=0.8\columnwidth]{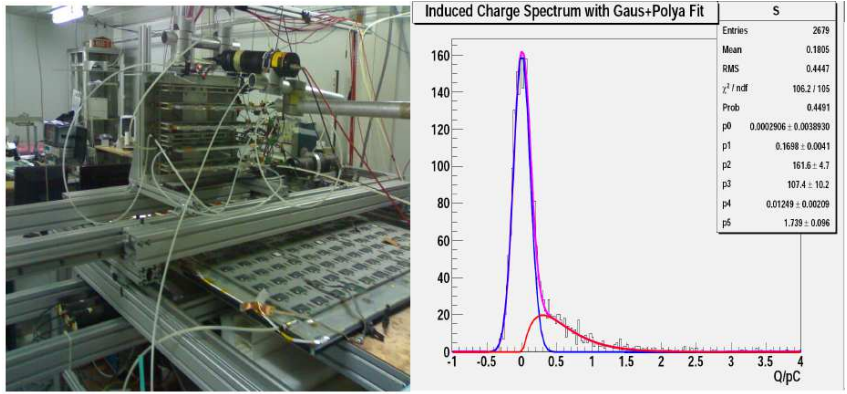}
\caption{Cosmic ray experiment and Induced charge spectrum}
\label{cosmicray}
%\endcenter
\end{figure}

%\begin{figure}
%\centering
%\includegraphics[width=0.6\columnwidth]{Correlation.eps}
%\caption{Correlation between induced charge and deopsited energy for 100GeV Pion}
%\label{QandE}
%\end{figure}

\section{Digitization module for GRPC SDHCAL}

The digitization module is important for the GRPC SDHCAL option since people are eager to know if the SDHCAL option could successfully record more useful information than DHCAL. 

As shown in Fig.~\ref{cosmicray}, a cosmic ray experiment has been performanced. Using the scintillator as trigger, the anaylog readout of one channel has been recorded, and reconstruced into the induced charge spectrum with about 2700 cosmic ray events.
The induced charge spectrum could be fit to a Polya function as signal and a Gaussian pedestal noise. 
The fitted parameters of the spectrum are used as the input for the digitization.

The method of digitization is straightforward. 
First, readout the Monte Carlo truth information of energy deposition from simulation;
second, estimate the number of ionizations that will be generated according to the energy deposition. 
The induced charge to one individual ionization is estimated from a Polya function (with 
the parameters calculated from the experimental input).
And the total induced charge is assumed to be sum of the charge of all the ionization. 

\begin{figure}[h]
\centering
\centerline{\includegraphics[width=0.8\columnwidth]{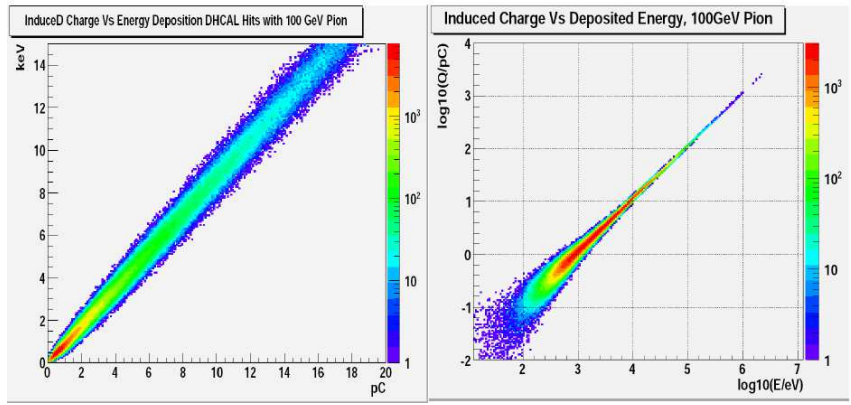}}
\caption{Correlation between induced charge and deopsited energy for 100GeV Pion}
\label{QandE}
\end{figure}

The correlation of induced charge and deposited energy is shown in Fig.~\ref{QandE}. Because of large uncertaintly of the RPC avanache development,
there is large smearing in charge inducing at low energy deposition region, but statistically nice correlation has been observed at high energy deposition region, where the induced charge could successfully represnent the energy deposition information. In other word, SDHCAL do keeps more energy information comparing to the pure DHCAL. 

The current digitization module has not taken into account the multiplicity and saturation effects. 
Those corrections will be included in a upgraded version.

\section{DRUID: Displaying Root module used for ILD}

\begin{figure}
\centering
 \includegraphics[width=0.46\columnwidth]{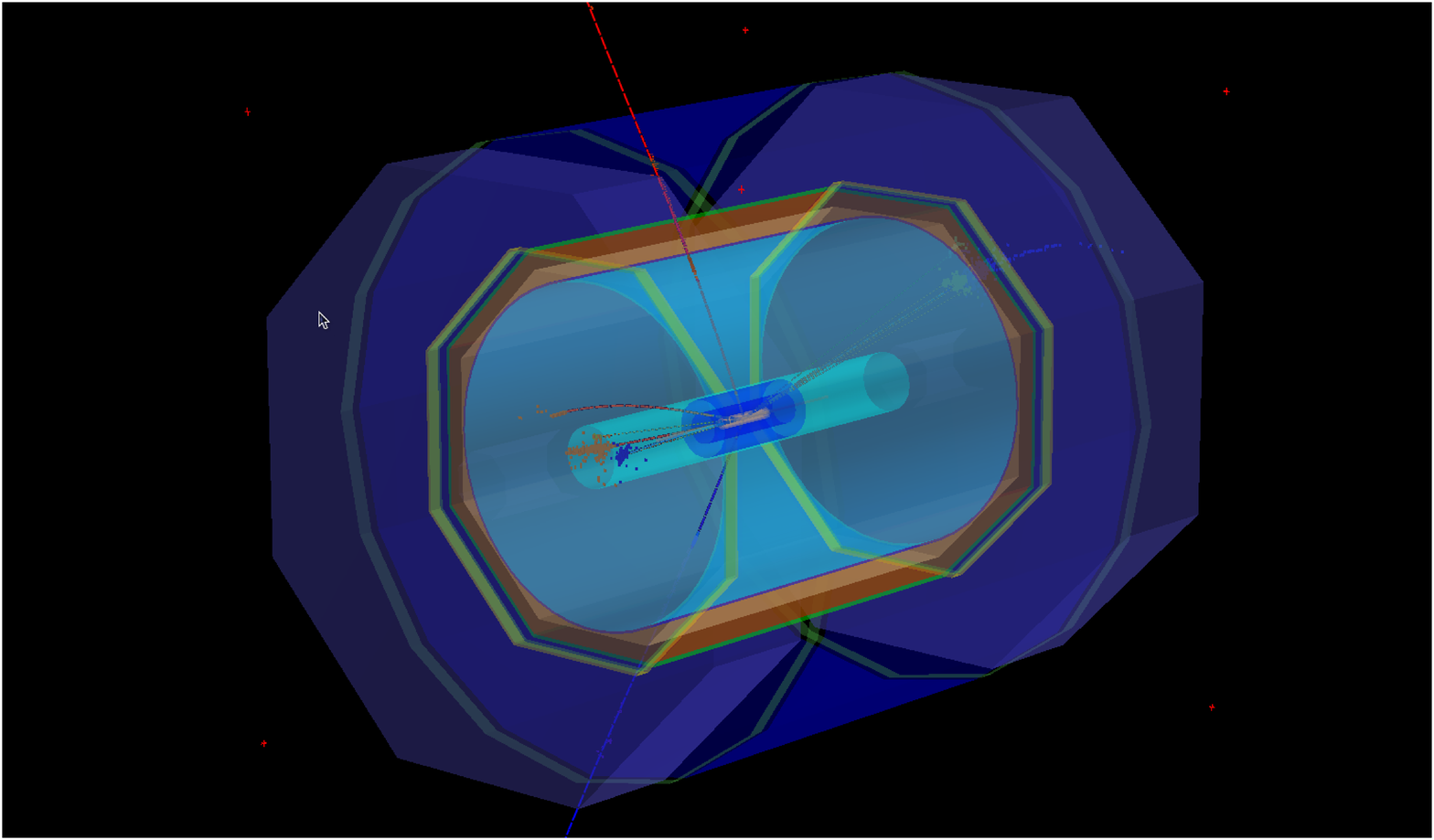}
\hspace{0.1in}
 \includegraphics[width=0.47\columnwidth]{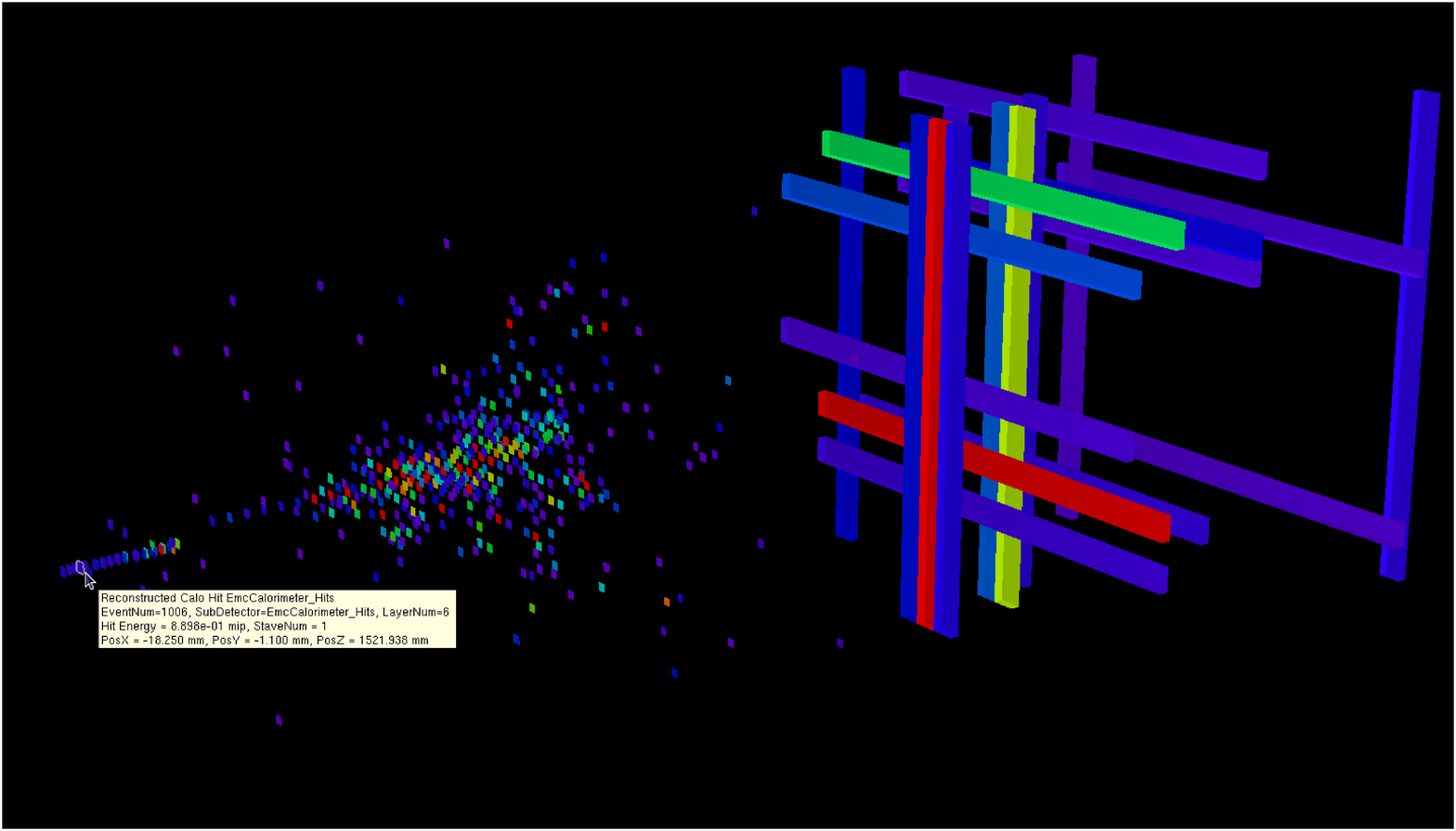}
\vspace{0.2in}
\caption{\textbf{Left:} Simulated ZH event at ILD detector \textbf{Right:} Reconstructed CALICE test beam event}
\label{Calice}
\end{figure}

Druid \cite{DruidNote} is a 3d display program used to visualize information stored in standard ilc data file (slcio format). 
Currently it is specified to support the ILD geometries with gear file as geometry input file, 
and has been tested on simulated ILD event and CALICE test beam event (Fig.~\ref{Calice}). 
According to different collections stored in the lcio file, Druid could display different objects in different combinations and styles. 

Druid needs a pre installed ROOT v5.22.00, GEAR v0.12.0 and LCIO v1.11 (or higher).
%Those could be accessed at the root homepage (\url{http://root.cern.ch}) and the ilcsoft portal (\url{http://ilcsoft.desy.de/portal}).
It can be accessed at IN2P3 svn server %\url{https://llrforge.in2p3.fr/svn/Druid} 
or my personal webpage \\ \url{http://llr.in2p3.fr/~ruan/ILDDisplay} with a brief manual.
%  \url{http://llr.in2p3.fr/~ruan/ILDDisplay/DruidNote.pdf}.

\subsection{Objects to be displayed}

\subsubsection{Detector geometry}

Currently Druid support the ILD full detector geometry with a la Videau HCAL or TESLA HCAL, and the frame work for CALICE test beam geometry. 
Different subdetectors could be easily mounted and dismounted by interactivly clicking on the GUI.

%For the ILD full detector, only subdetectors inside the HCAL are displayed. So don't be surprised if you see some Muon detector hits outside the displayed detector.

\subsubsection{Event information}

Event information is organized into different groups:

\vspace{.5cm}

MCParticle: MCParticles are displayed as tracks or arrows. 

Calorimeter Hit: simulated/digitized calorimeter hits, displayed as cuboids with different size, orientations and color.
%(according to different subdetector) and colors (according to energy, PID of the particle that creates this hit, PID of the origin of this hit (mother particle at the vertex), randomly generated color and uniform color ).

Tracker Hit: simulated/digitized tracker hits, displayed as points.

%Calorimeter Hit: Digitized calorimeter hits, as well as simulated calorimeter hits, 
%we have different orientation and different colors (according to energy information) available for each cuboid.

%Tracker Hits: Digitized tracker hits.

ReconstructedParticle: reconstructed particles are displayed as tracks with the calorimeter hits assigned to this particle.

% as PID of reconstructed particle, 
%randomly generated color and uniform color.

\vspace{.5cm}

Here MCParticle and reconstructed particle collection are divided into subgroups according to their PID/energy, 
and the hits groups are divided into subgroups according to the subdetectors. 

%For the MCParticle, there exist a subgroup called 'Mothers', which record the information of Mother particles at the Vertex, indicating the event type. 
%The mother particles are displayed as arrows. 
%Each simulated calorimeter hit is also assigned with the mother particle information: they could be colored according to the mother particle index.

Different groups of objects could be hidden/displayed. 
Druid will remember the status of display/hide for different objects from last event, and keep the status when navigated to a new event. 
By default, Druid will always display the new object. 

%That's to say, Druid will display every object for the first event (Only one exception: the LowE, NeutralHad and Neutrons subgroup of MCParticle, corresponding to the collection of low energy particle, neutral hadrons and neutrons, will not be displayed by default. Since they are regarded as event detail and their huge number will make the display too mess). 

%\subsection{Reconstructed information}

%\section{Display options}

%\begin{wrapfigure}{r}{0.27\columnwidth}
%\begin{figure}
%\centerline{\includegraphics[width=0.25\columnwidth]{Options.eps}}
%\caption{Druid Option Pallel}
%\label{option}
%\end{wrapfigure}

%\section{Display options}

\subsection{Display options}

Druid display could be easily zoomed, rotated and shifted by the mouse. 
With a self defined GUI pannel, lots of interactive actions are supported in DRUID. 
%For example, navigation to different event either by their order (go to previous/next event) or by giving directly the event number.

%\subsubsection{Event Navigation}
%Two ways of event navigation has been provided in the GUI page of 'options'. 
%By click on the button with forward/backward arrow, you could navigate to next/previous event; 
%or you can enter a specified event number into the box 'Go to Evt', and then type enter:
%Just make sure you have the right event number.

\subsubsection{Individual objects}

TEve allows us to attach text information on individual objects, which could be readout by picking up the object with the mouse. 

\begin{figure}
\centerline{\includegraphics[width=1.0\columnwidth]{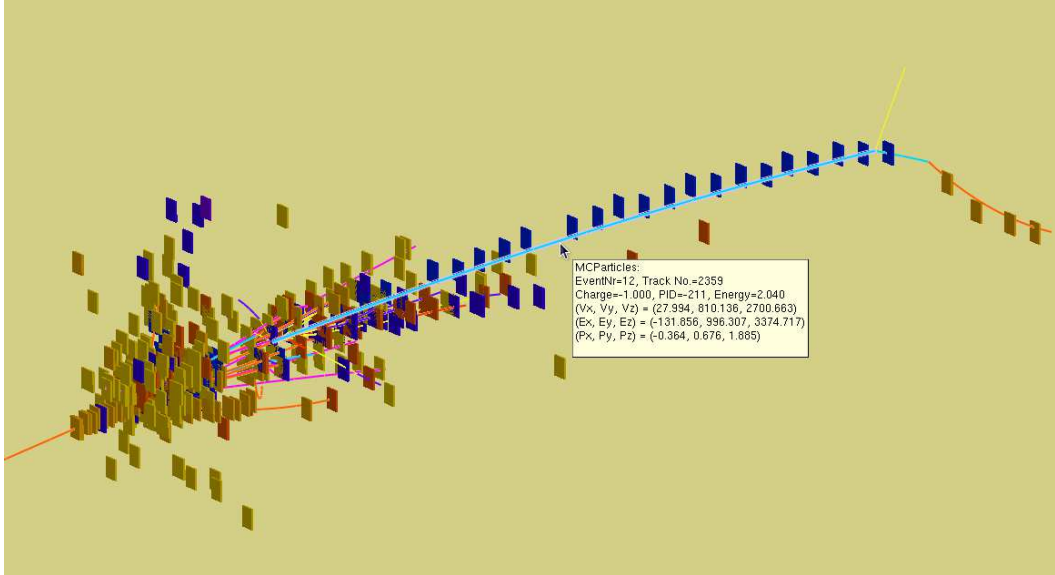}}
\caption{40 GeV pion shower in the ILD Calorimetry}
\label{40GeVPion}
\end{figure}

The rotation center could be selected, together with the zoom option, you can have a very closed and clear view of event details. 
Fig.~\ref{40GeVPion} shows the shower created by 40GeV Pion, with the Monte Carlo truth information of the secondary particles generated inside the hardonic shower. 
Where you can see a 2.04GeV Pion is generated in the interaction, and pass through the HCAL as a mip. 

\subsubsection{Option for calorimeter hits}

Druid allows various of options for the size and color of calorimeter hits, which could be specified in the option pannel. 
For example, simulated calorimeter hit could be coloured according to the hit energy, the PDG of MCParticle that passing through or at the vertex (as origin), or a randomly generated color. As for the reconstructed particle assigned calorimeter hit, it could be coloured to the PDG information of reconstructed particle.

\begin{figure}
\centerline{\includegraphics[width=0.9\columnwidth]{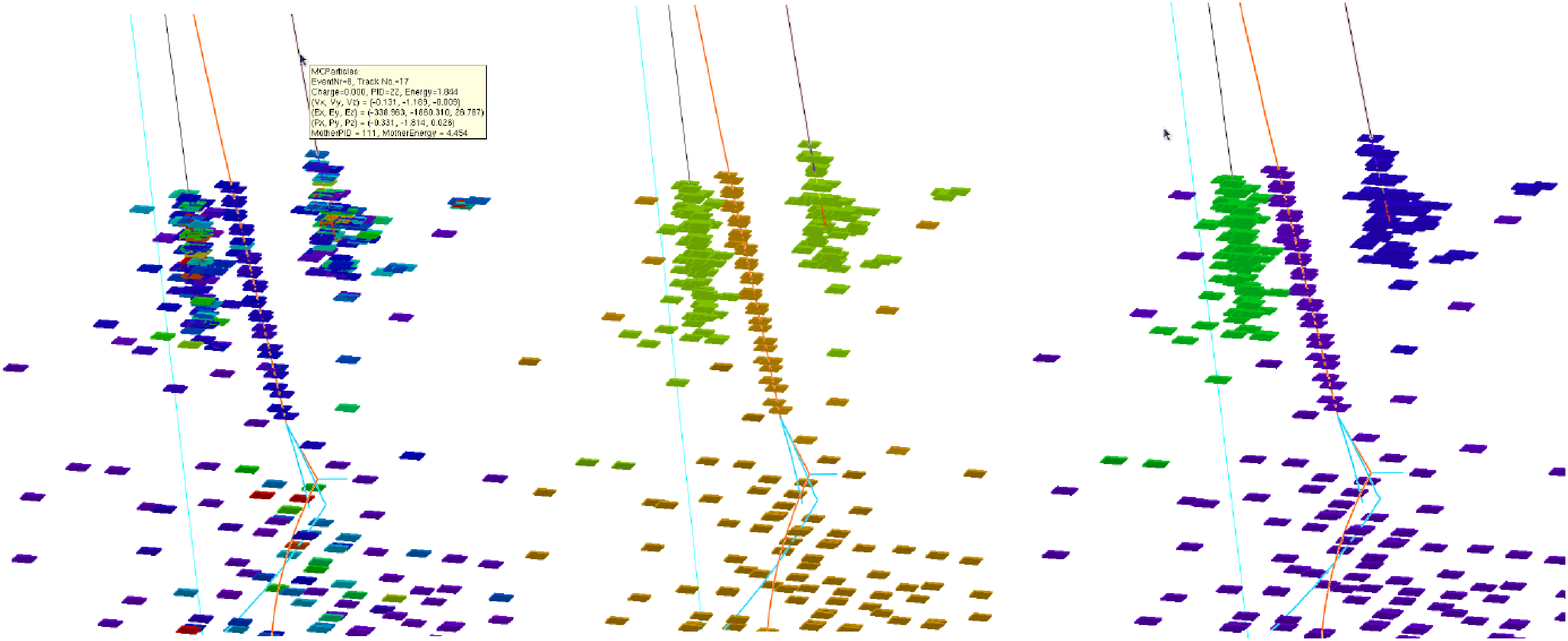}}
\caption{Tau jet $\tau \rightarrow \nu + \pi^{0} + \pi^{+}$ with different color options: from left to right, energy, track ID and random index on track}
\label{coloroption}
\end{figure}

Fig.~\ref{coloroption} shows how a $\tau$ jet looks like with different color option (at MC truth level).

\begin{figure}
\centerline{\includegraphics[width=0.9\columnwidth]{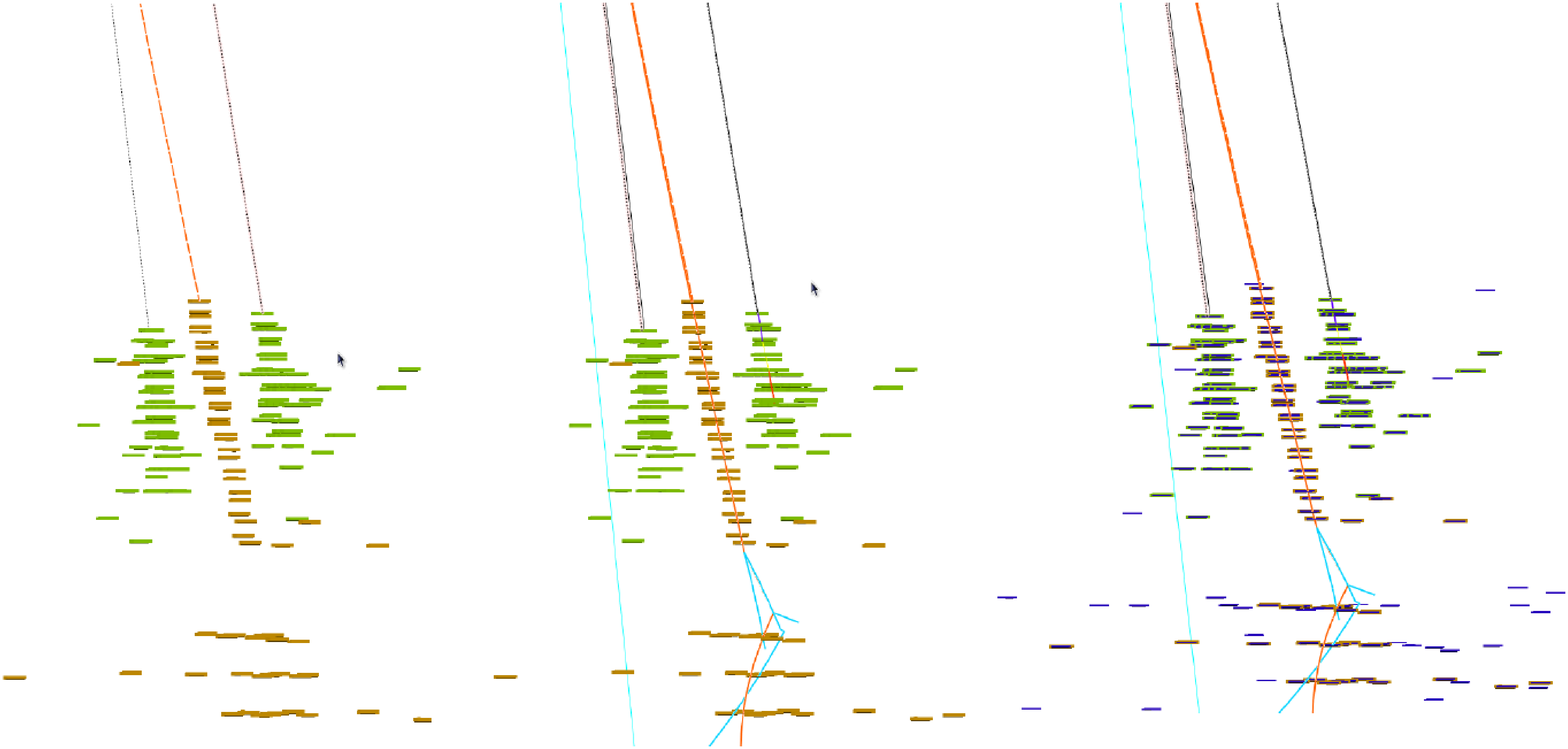}}
\caption{Tau jet $\tau \rightarrow \nu + \pi^{0} + \pi^{+}$ at MCTruth and Reconstruction level}
\label{Reccompare}
\end{figure}

%For the color option with energy, we use standard root palette 1, and the default setting is to make blue for 1 mip hits, 
%red for hits energy larger than 10 mip and purple for hits energy smaller than 1 mip. 

%For user interested in low energy hits, we provided a global scale factor (with initial value = 10) function to change the color setting 
%If you want to analysis the high/low energy hits, than reduce/increase this factor; 
%(i.e, set Cell Color Scale = 1 means 10 mip hit will have blue color, and 100 mip hit have blue color, etc).

%An additional option for HCAL hits according to the semi-digital HCAL is provided, 
%where the hits color is assigned to 4 kinds with the thresholds information. 

%where you can specify the three thresholds (with ratio 1:10:100) between each. When this option is active (with setting thresholds larger than 0), we have purple for hits with energy lower than the smallest threshold, blue for hits with energy between 1st and 2nd threshold, green for hits betweern 2nd and 3rd threshold, and red for hits with energy higher than largest threshold. 

%Each time you change the color or size option, the screen will output the list of collections that has been redraw.

\subsection{Analysis reconstruction algorithm performance}

By comparing the Monte Carlo truth objects and the digitized/reconstructed objects, DRUID provides a platform to analysis the reconstruction software performance.  

Fig.~\ref{Reccompare} shows an example with the same $\tau$ jet at MC truth and reconstruction level. 
The left one is the reconstructed particle tracks (using PandoraPFA) with assigned calorimeter hits, where you can see Pandora found two gamma and one pion;
The middle one overlay the MCParticle information, thus we know that Pandora is right about the gammas and pions, 
also some secondary particles generated inside the shower and the neutrino could be found. 
The right one overlay the simulated calorimeter hits: some simulated calorimeter hits are dropped during the digitization/reconstruction. 
You can easily tell if this dropping is expected or not by reading the text information attached to each of those hits.  

\subsection{Conclusion}

%GRPC semi-DHCAL technology is a solid option for the ILD, and people of the GRPC semi-DHCAL group are making efforts both on the hardware development and the software developments. 

As a solid option for the ILC, the software development of GRPC SDHCAL technology is progressing at steady pace with the hardware developments. 
%To prove the GRPC SDHCAL option is feasible for the ILD detector, people are making efforts on various directions. 

The first order of digitization module has been developed, and the preliminary result shows that SDHCAL can record more information than the DHCAL, 
ensures better performance. 

The display package DRUID has been developed, with which people could get better understanding to the ILD/test beam event as well as to the performance of reconstruction algorithms. 

The digitization module will be upgraded with realistic physics efforts (saturation, efficiency and multiplicity) and the DRUID will be updated with new geometries. As a central task for Semi DHCAL software development, people are currently foucs on the design of reconstruction algorithm, which will be integrated into the PandoraPFA reconstruction framework.

%\section{how to play}

%% section headers !

%\section{Acknowledgments}

\section{Bibliography}

% ****************************************************************************
% BIBLIOGRAPHY AREA
% ****************************************************************************

\begin{footnotesize}
% IF YOU DO NOT USE BIBTEX, USE THE FOLLOWING SAMPLE SCHEME FOR THE REFERENCES
% ----------------------------------------------------------------------------

% ----------------------------------------------------------------------------

\end{footnotesize}

% ****************************************************************************
% END OF BIBLIOGRAPHY AREA
% ****************************************************************************


\begin{thebibliography}{99}
% Please replace the numbers for   contribId   and   sessionId
% in the following URL. You can get this information by going to 
% http://indico.cern.ch/confAuthorIndex.py?confId=2628
% and search for your contribution and click on the title
% Be aware: '&amp;' must be replaced by simple '&' as in example below
%------- replace following references ;-)

\bibitem{Pandora} \url{http://www.hep.phy.cam.ac.uk/~thomson/pandoraPFA}
\bibitem{rootpage} \url{http://root.cern.ch}
\bibitem{DruidNote} \url{http://llr.in2p3.fr/~ruan/ILDDisplay/DruidNote.pdf}
%\bibitem{lcwspresentation} \url{http://llr.in2p3.fr/~ruan/ILDDisplay}

%M. Ruan, Semi-DHCAL software development: Digitization $\&$ Display, presentation at LCWS 2010, Beijing

%\bibitem{parton_qed} A.D.~Martin {\it et~al.}, Eur. Phys. J. {\bf C39} 155 (2005).
%\bibitem{H1}N.~Gogitidze, arXiv:hep-ex/0701033 (2007).
%\bibitem{DVCS}S.~Friot, B.~Pire and L.~Szymanowski, Phys. Lett. {\bf B645} 153 (2007);\\
%              D.~Hasell, R.~Milner and K.~Takase, AIP Conf. Proc. {\bf 588} 187 (2001);\\
%              M.~Krawczyk and A.~Zembrzuski, Phys. Rev. {\bf D57} 10 (1998).
%\bibitem{pomeron}R.~Brower and C.~Tan, PoS LAT2005 279 (2006);\\
%                 J.P.~Guillaud and A.~Sobol,
%  {\it Perspectives of the study of double Pomeron exchange at the LHC},
%  11th Lomonosov Conference on Elementary Particle Physics, Moscow, Russia (2003).
\end{thebibliography}
\end{document}